\documentclass[aps,prl,twocolumn,groupedaddress]{revtex4}
\usepackage{epsfig}
\begin{document}

\title{Chiral susceptibility in canonical spin glass and reentrant alloys from Hall effect measurements}
\author{ P.Pureur, F. Wolff Fabris , J. Schaf }
\affiliation{Instituto de F\'{i}sica, Universidade Federal do Rio Grande do Sul, \\
Caixa Postal 15051, 91501–970 Porto Alegre, RS, Brazil}
\author{I.A. Campbell}
\affiliation{Laboratoire des Verres, \\ Universit\'e Montpellier II,
       34095 Montpellier, France}


\begin{abstract}

The extraordinary Hall effect coefficients $R_s(T)$ of the canonical spin glass alloys ${\bf Au}Fe 8 at\%$ and ${\bf Au}Mn  8 at\%$ and of an archetype reentrant system ${\bf Au}Fe 18 at\%$ were measured as functions of temperature. The data show a critical cusp-like term superimposed on a smooth background for the spin glasses, and change of sign with temperature for the reentrant. The results can be interpreted consistently by invoking a chiral Hall effect contribution as proposed by Kawamura.
\end{abstract}


\maketitle

Spin glasses have been intensively studied for more than thirty years, as paradigms for the
statistical mechanics of the whole vast family of complex systems. However the origin the spin freezing in the canonical spin glasses, dilute alloys such as ${\bf Au}Fe$ or ${\bf Cu}Mn$ where the spins are Heisenberg,  has long remained an enigma. Experiments and in particular critical exponent measurements \cite{exponents} have shown definitive evidence for a non-zero ordering temperature, while numerical work on Heisenberg spin glasses in dimension three indicated that there was an Edwards-Anderson type of ordering only at zero temperature \cite{Tg0} though recent simulations contradict this \cite{nakamura:02}. For vector spin glasses there exists a chiral order parameter in addition to the Edwards-Anderson spin parameter and there were early suggestions that chirality might play a role in vector spin glass ordering \cite{villain:77}. Kawamura and coworkers \cite{kawamura:92,hukushima:00,imagawa:03a} have made concrete large scale numerical investigations of the chiral driven ordering mechanism which they postulated to explain freezing in Heisenberg spin glasses. In real samples the spin and chiral order are linked through Dzaloshinsky-Moriya (DM) random anisotropy terms which are inevitably present. Magnetic torque experiments on a range of spin glasses have shown in-field transition lines up to high applied fields \cite{petit:99}, in excellent agreement with extensive simulations on Heisenberg systems submitted simultaneously to anisotropy and field, where both chiral and spin ordering were monitored \cite{hukushima:00,imagawa:03b}. The robustness of order under applied fields is a clear indication that chirality is a primary ingredient of spin glass ordering in vector systems. The torque measurements were interpreted using the chiral ordering scenario. 

In this context it would be of considerable interest to have a complementary direct observation of chirality in the spin glasses. Chirality is a "hidden" parameter and no technique was known with which to monitor it experimentally, until Kawamura proposed that the Hall effect should present a direct signature of the chiral susceptibility \cite{kawamura:03}. On the chiral scenario the extraordinary Hall signal (linked to the sample magnetization rather than the magnetic field) should have a critical term due to chirality. The key point is whether there is critical behaviour of the Hall coefficient $R_s(T)$ at the glass temperature $T_g$, and accompanying  non-linear effects. $R_s$ is defined as the ratio of the Hall signal to the magnetization, once corrections have been made for the ordinary Hall term. Careful Hall experiments on spin glasses have been reported \cite{mcalister:76,barnard:88} but at the time of the measurements no appropriate theoretical framework had been proposed with which the data could be interpreted. We have studied two canonical spin glasses, ${\bf Au}Fe 8 at\%$ and ${\bf Au}Mn 8 at\%$, and an archetype reentrant system ${\bf Au}Fe 18 at\%$. We observe behaviour for $R_s(T)$ clearly indicating a chiral contribution as predicted by Kawamura. These experiments can be considered to be "direct" observations of chiral susceptibility in canonical spin glasses. 

Ingots of the different alloys were prepared by arc melting the constituents under an atmosphere of pure argon. Purities of the starting metals were $99.9985\%$ for Au and $99.99\%$ for Fe and Mn. Each ingot was rolled to about $0.5$ mm and a bar shaped sample was cut out for the magnetization measurements. The remaining material was further rolled to $25 \mu m$ in the case of ${\bf Au}Fe$, and to $10 \mu m$ , in the case of ${\bf Au}Mn$. From these foils, samples of $10 mm$ x $2 mm$ with 5 projecting contacts were prepared for resistivity and Hall experiments. Each couple of samples of each alloy was sealed in a quartz ampoule under argon and  annealed at $900^{\circ} C$ for $1 h$ before quenching into water. 
	A Quantum Design MPMS-XL SQUID magnetometer was used for the magnetic moment measurements. Experiments were performed as functions of the temperature in fixed fields according to the ZFC and FC protocols. The applied fields used ranged from   
$\mu_{0}H = 0.015 T$ to $2 T$. $T_g$ values are $28 K$ and $24 K$ respectively for the ${\bf Au}Fe 8 at \%$ and ${\bf Au}Mn 8 at \%$ samples. ${\bf Au}Fe$ has moderately strong DM anisotropy \cite{petit:99} while there appears to be no information published on the anisotropy of ${\bf Au}Mn$.

	The longitudinal and Hall resistivities were measured with an AC technique operating at $37 Hz$. For the Hall measurements, the signal associated with the longitudinal resistivity was used in a compensation circuit in order to eliminate spurious magnetoresistance contributions. The temperatures were determined in the tranport experiments with a Lake-Shore calibrated carbon-glass thermometer. An RMS current of $30 mA$ was applied in the Hall experiments, which were performed according both the ZFC and FC protocols. In the ZFC measurements, the sample was cooled in the absence of field to some fixed temperature below $T_g$. Then the Hall voltage was recorded while the field was increased in the same order as for the magnetization measurements. After each run sequence, the field was again set to zero and the sample was heated up to a temperature well above $T_g$. The cycle re-starts by cooling the sample down to another temperature point. In the FC experiments, the field was kept constant while ramping the sample temperature down gradually to the lowest temperature studied. Since the Hall voltages are quite small in these noble metal spin glasses, special care must be taken in order to subtract the background signals, specially in the FC experiments. Below $T_g$ spin glass magnetization is very sensitive to field cycling effects; in the present context it is important to carry out measurements of Hall effect and magnetization under identical experimental protocols. Ideally, in order to probe the initial linear regime Hall experiments should be performed in a very low applied field. In practice with the experimental sensitivity available the minimum field for which exploitable Hall data could be obtained was $0.015 T$. The experiments were repeated at higher fields also so as to obtain information to be used for identifying the side-jump term and to explore the non-linear behaviour. The relative accuracy of the magnetization is much higher than that of the Hall signal.  

In magnetic field $B$ a sample with magnetization $\mu_{0}M$ shows a Hall resistivity defined through
\begin{equation}
\rho_h = \mu_{0}R_{h}H = R_{0}\mu_{0}H +R_{s}\mu_{0}M
\end{equation}
$R_0$ is the "ordinary" Hall coefficient which leads to a term which should vary little with $T$. Conventionally $R_{s}\mu_{0}M$ is known as the "extraordinary" term. Extrapolations to high temperatures to obtain an estimate of the zero magnetization contribution for the present alloys, and for various other Au based spin glass samples \cite{barnard:88}, indicate that for all the dilute alloys $R_0$ is close to $-7 .10^{-11} m^{3}C^{-1}$; this is equal to the "low field" value for the pure metal \cite{hurd:72}. In the region of $T_g$ for the present alloys the corresponding ordinary contribution to $R_h$ is small compared to the magnetic contribution; it has been assumed that $R_0$ is temperature independent in all the data analysis.  In the magnetization measurements the sample is in the form of a thin plate with the field parallel to the plane so the demagnetization factor is negligible. For the Hall geometry on the other hand the demagnetization factor is necessarily maximum, and when comparing the two sets of data to estimate $R_s(H,T)$ it is simple but essential to take the demagnetization correction into account \cite{barnard:88}. 

Kawamura \cite{kawamura:03} showed that in samples with canted spins one can expect  a chiral Hall term proportional to the chiral susceptibility. He predicted a chirality-susceptibility linked cusp-like anomaly in $R_s(T)$ at $T=T_g$ in the low field limit linear regime, possibly accompanied  by the onset of a deviation between FC and ZFC curves. (It is important to note that the predictions concern the coefficient $R_s(T)$ not the Hall signal $\rho_h(T)$ itself, which will have a singularity at $T_g$ just because $M(T)$ has a singularity). The singularity would be expected to round off at higher fields because of non-linear effects, as is the case for the susceptibility.  The strength of the chiral Hall signal can be expected to be linked to the spin-orbit coupling which also gives rise to the DM anisotropy, but neither the sign nor the strength of the predicted term linked to the chirality can be predicted simply as they depend in a complex manner on the electronic structure of both the host material and of the magnetic sites. There will also be standard terms contributing to $R_s$ known as the skew-scattering and the side-jump effects which have been studied in detail in conventional ferromagnetic alloys \cite{skewsidejump}. They too are consequences of spin-orbit coupling, but should contribute a smoothly changing background term to $R_s(H,T)$ without any singular behaviour at $T_g$. Both depend on the sample resisitivity $\rho(T)$; in high resisitivity samples like those studied in the present work the dominating standard term is the side-jump for which $R_s(T)$ is proportional to $\rho^{2}(T)$. $\rho(T)$  changes smoothly and only weakly over the range of temperature and field used. 

\begin{figure}[h]
\resizebox{0.4\textwidth}{!}{%
\includegraphics[scale=0.2]{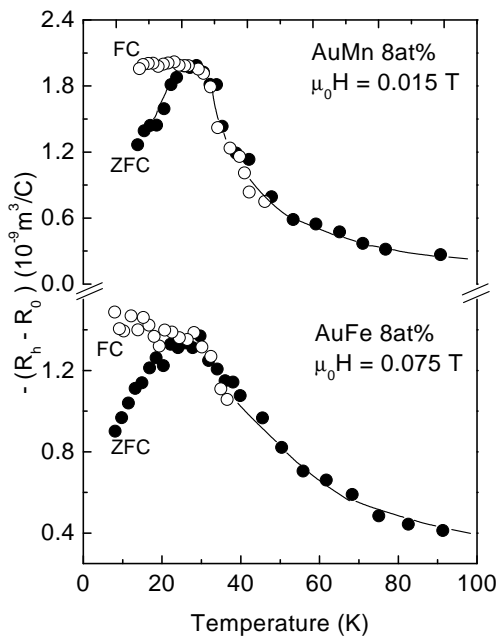}
}
\caption{The Hall effect corrected for the ordinary term $[R_h(T)-R_0]$ for ${\bf Au}8\% Mn$ and ${\bf Au}8\% Fe$ under FC and ZFC protocols.}
\label{Figure:1} 
\end{figure}

Figure 1 shows $R_{h}(T)-R_{0}$ data corresponding to the FC and ZFC protocols  as functions of temperature for the two spin glass samples. In both cases there is a cusp in the ZFC $R_h(T)$ near $T_g$, accompanied by a splitting between FC and ZFC curves which develops just below $T_g$. 
 We transpose the data into the form of $R_{s}(T)$ plots by dividing $[R_{h}(T)-R_{0}]B$ by $\mu_0M_{h}(T)$, where $\mu_0M_{h}(T)$ is the magnetization measured in the SQUID $\mu_0M(T)$ corrected for the demagnetization factor, $M_{h}(T)=M(T)/(1+M(T))$. The data for $R_s(T)$ under the FC and ZFC protocols for the ${\bf Au}Fe$ and ${\bf Au}Mn$ spin glasses at a field $\mu_0H = 0.015 T$ are shown in Figure 2. Remarkably the absolute value of $R_s(T)$ is minimum near $T_g$. The only plausible interpretation would appear to be in terms of two competing contributions to the total signal : a negative background term having a smooth non-critical temperature dependence, and superimposed on it a weaker positive term with a critical maximum near $T_g$. 
To analyse the data we will assume that the background term in $R_s(T)$ is dominated by the side-jump contribution and that it remains equal to $\rho^2(T)$ multiplied by a constant over the entire $T$ range. The value of the constant is estimated by measuring to temperatures well above $T_g$ and extrapolating to infinite $T$. The estimated side-jump contribution $R_s^{sj}$ for the ${\bf Au}Fe$ sample is shown in Figure 2; the ${\bf Au}Mn$ side-jump term is of very similar form. Subtracting out the side-jump term obtained this way from the total $R_s(T)$ leaves us with a positive difference term. This procedure although necessarily only approximate is the most satisfactory available.

Quoting all $R_s$ values in units of $10^{-8}m^{3}C^{-1}$, the side-jump contribution is $R_s^{sj}\approx -4.8$ in the region near $T_g$ for the ${\bf Au}Fe$ sample, and  $\approx -1.9$ in the same region for the ${\bf Au}Mn$ sample. The difference term $[R_s - R_s^{sj}]$ peaks in the region of $T_g$ at a value of $\approx +2.8$ in the ${\bf Au}Fe$ sample and at $\approx +1.1$ for the ${\bf Au}Mn$ sample. In both cases the FC and ZFC $R_s(T)$ difference curves separate below $T_g$. For both alloys extrapolation suggests that the difference term at these low fields tends to zero or close to zero as $T$ tends to zero or to high $T$.

\begin{figure}[h]
\resizebox{0.4\textwidth}{!}{%
\includegraphics[scale=0.2]{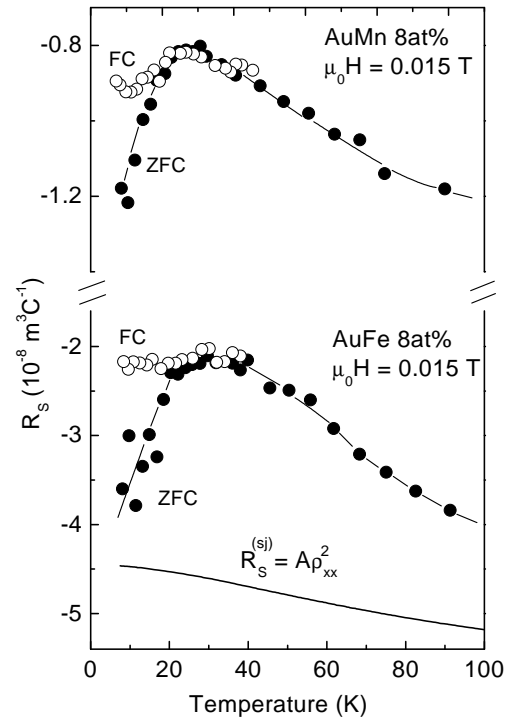}
}
\caption{The extraordinary Hall coefficient $R_s(T)$ for ${\bf Au}8\% Mn$ and ${\bf Au}8\% 
Fe$ under FC and ZFC protocols. The line indicates the side-jump contribution $R_{s}^{sj}(T)$ for the ${\bf Au}Fe$ sample estimated from high temperature measurements.} 
\label{Figure:2} 
\end{figure}

We identify the difference term with the chiral term proposed by Kawamura \cite{kawamura:03}; no other Hall effect model has been put forward which predicts a contribution with a critical cusp in $R_s(T)$ at $T_g$, accompanied by separation between FC and ZFC $R_s(T)$ curves below $T_g$. Unfortunately, in the present work the lowest fields which could be used do not provide us with strictly low field limit data to test the detailed predictions of \cite{kawamura:03}. However the present data suggest that the chiral term cusp would not be sharp even in the low field linear limit before corruption by non-linear effects. It should be noted that the data reported by \cite{barnard:88} on nominally the same concentration ${\bf Au}Mn$ alloy as in the present work but using much lower measuring fields would also show a flat topped cusp extending somewhat above $T_g$ if plotted as in Figure 2. 

The analysis of the non-linear behaviour of $R_s(T,H)$ will be discussed elsewhere.

\begin{figure}[h]
\resizebox{0.5\textwidth}{!}{%
\includegraphics[scale=0.5]{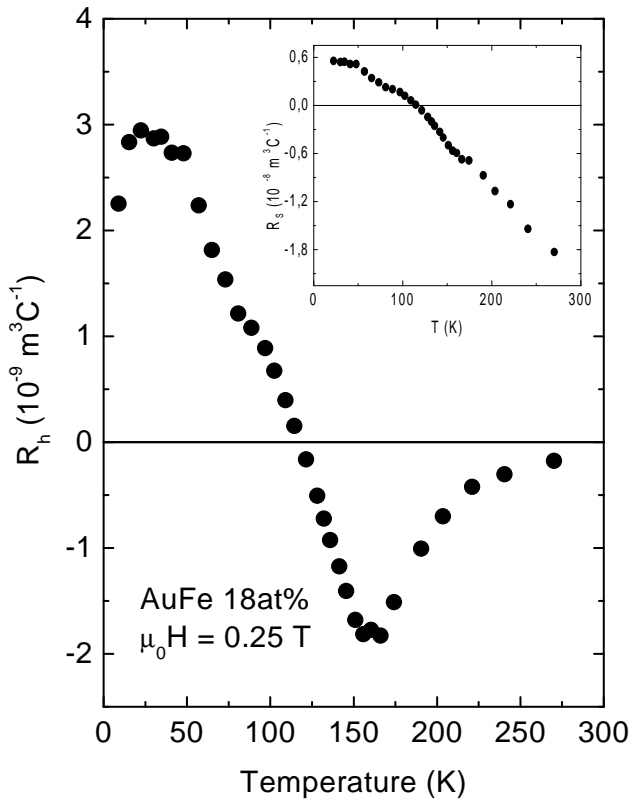}
}
\caption{The Hall coefficient $[R_h(T)-R_0]$ for ${\bf Au}18\% Fe$ under an applied field of $0.25T$. FC and ZFC protocols give the same signal for this field. The inset shows the extraordinary Hall coefficient $R_s(T)$.}
\label{Figure:3} 
\end{figure}

Many properties of the archetype reentrant alloy ${\bf Au}Fe 18 at\%$ have already been investigated experimentally in great detail \cite{reentrant}. At this concentration the alloy has a Curie temperature near $165K$, and a canting temperature $T_k$ near $75K$ below which transverse components of the local moments freeze while the overall magnetization is practically conserved \cite{reentrant}. Static chiral properties should appear in reentrant systems in the region of $T_k$ when transverse moments develop \cite{kawamura:03} (data on more exotic reentrant systems have been tentatively interpreted invoking a chiral term \cite{aito:03}). As in the spin glasses, one can expect a chiral susceptibility term already above $T_k$ and it should be remembered that because the Hall effect is a transport measurement, short time scales are implicity involved.  At temperatures close to but above $T_k$ there will be rapid transverse magnetization fluctuations, and these are indeed observed in neutron scattering \cite{hennion:95}.

For the ${\bf Au}Fe 18 at\%$ sample, the Hall coefficient $R_h(T)$ in a field of $0.25T$ is shown in Figure 3. The magnetization for a ferromagnet in the Hall geometry is practically equal to $\mu_0H$ (until $\mu_0M  \geq \mu_0H$) and so independent of $T$ below $T_c$.   In contrast, we find a Hall coefficient $R_h(T)$ which varies dramatically with temperature in this sample. It begins negative in the paramagnetic regime and in the high temperature ferromagnetic region, before changing sign to positive near $T \sim 110K$. As in the spin glass alloys, the temperature dependence can again be understood in terms of competing negative side-jump and positive chiral terms. The former is the only term present when the temperature is high so the time average local moments are all parallel to the total magnetization and therefore there is no chirality; the chiral term should appear as $T$ drops and the local moments become canted with respect to the global magnetization allowing chirality. It turns out that in this sample the chiral term dominates for temperatures below about $110K$ explaining the sign change. This scenario with a smoothly varying negative side-jump term and a positive chiral term taking over when chirality is permitted is a satisfying confirmation of the interpretation given above for the  spin glass data. The data expressed in terms of $R_s(T)$ are shown in the inset of Figure 3.

In conclusion, confirming predictions made by Kawamura \cite{kawamura:03}, the extraordinary Hall coefficient of canonical spin glass and reentrant systems shows a contribution linked to the onset of ordering in the spin glasses, and with canting in the reentrant alloy. These results can be analysed to provide a direct signature of the chiral susceptibility \cite{kawamura:03} in these systems. It is the first time that chiral susceptibility, a fundamental property of all magnetic systems with canted spins, has been identified directly in canonical Heisenberg spin glasses rather than implicitly through its influence on the spin order.

\end{document}